\documentclass[sigconf]{acmart}

\usepackage[utf8]{inputenc}
\usepackage[T1]{fontenc}

\usepackage{siunitx} 

\usepackage{tikz} 

\newcommand\circled[1]{{\textcircled{\hspace{0pt}\scriptsize{\raisebox{0.2pt}{#1}}}}}

\usepackage{pifont}

\newcommand{\new}[1]{#1}

\usepackage{xspace}
\newcommand{\tool}{\textit{Test\-Spark}\xspace}
\newcommand{\evosuite}{\textit{EvoSuite}\xspace}

\newcommand{\idea}{IntelliJ IDEA\xspace}
\newcommand{\grazie}{an internal service API in JetBrains\xspace}
\newcommand{\openai}{\textit{OpenAI}\xspace}
\newcommand{\chatgpt}{\textit{ChatGPT}\xspace}
\newcommand{\uut}{\textit{UUT}\xspace}

\newcommand{\eg}{\textit{e.g.,~}}

\newcommand{\ie}{\textit{i.e.,~}}

\newcommand{\etal}{\textit{et al.}\xspace}

\usepackage{listings}
\usepackage{xcolor}
\usepackage{todonotes}
\presetkeys{todonotes}{inline,color=yellow!40}{}

\definecolor{codegreen}{rgb}{0,0.6,0}
\definecolor{codegray}{rgb}{0.5,0.5,0.5}
\definecolor{codepurple}{rgb}{0.58,0,0.82}
\definecolor{backcolour}{rgb}{0.95,0.95,0.92}

\lstdefinestyle{mystyle}{
    backgroundcolor=\color{backcolour},   
    commentstyle=\color{blue},
    keywordstyle=\color{blue},
    numberstyle=\color{blue},
    stringstyle=\color{codepurple},
    basicstyle=\ttfamily\footnotesize,
    frame=single,
    breakatwhitespace=false,         
    breaklines=true,                 
    captionpos=b,                    
    keepspaces=false,                 
    numbers=none,                    
    numbersep=5pt,                  
    showspaces=false,                
    showstringspaces=false,
    showtabs=false,                  
    tabsize=2
}
\usepackage[nameinlink,capitalise]{cleveref} 

\AtBeginDocument{%
  \providecommand\BibTeX{{%
    \normalfont B\kern-0.5em{\scshape i\kern-0.25em b}\kern-0.8em\TeX}}}

\setcopyright{acmcopyright}
\copyrightyear{2024}
\acmYear{2024}
\acmDOI{XXXXXXX.XXXXXXX}

\acmConference[ICSE '24 Demonstrations]{46th International Conference on Software Engineering}{April 14--20,
  2024}{Lisbon, Portugal}
\acmPrice{15.00}
\acmISBN{978-1-4503-XXXX-X/18/06}

\begin{document}

 \title{TestSpark: IntelliJ IDEA's Ultimate Test Generation Companion}

\author{Arkadii Sapozhnikov}
\email{arkadii.sapozhnikov@jetbrains.com}
\orcid{0009-0002-1693-1937}
\affiliation{%
  \institution{JetBrains Research}
  \city{Berlin}
  \country{Germany}
}

\author{Mitchell Olsthoorn}
\email{M.J.G.Olsthoorn@tudelft.nl}
\orcid{0000-0003-0551-6690}
\affiliation{%
  \institution{Delft University of Technology}
  \city{Delft}
  \country{The Netherlands}
}

\author{Annibale Panichella}
\email{A.Panichella@tudelft.nl}
\orcid{0000-0002-7395-3588}
\affiliation{%
  \institution{Delft University of Technology}
  \city{Delft}
  \country{The Netherlands}
}

\author{Vladimir Kovalenko}
\email{vladimir.kovalenko@jetbrains.com}
\orcid{0000-0001-5880-7323}
\affiliation{%
  \institution{JetBrains Research}
  \city{Amsterdam}
  \country{The Netherlands}
}

\author{Pouria Derakhshanfar}
\email{pouria.derakhshanfar@jetbrains.com}
\orcid{0000-0003-3549-9019}
\affiliation{%
  \institution{JetBrains Research}
  \city{Amsterdam}
  \country{The Netherlands}
}

\renewcommand{\shortauthors}{Sapozhnikov \etal{}}

\begin{abstract}
    Writing software tests is laborious and time-consuming. 
To address this, prior studies introduced various automated test-generation techniques.
A well-explored research direction in this field is unit test generation, wherein artificial intelligence (AI) techniques create tests for a method/class under test.
While many of these techniques have primarily found applications in a research context, existing tools (\eg \evosuite, Randoop, and AthenaTest) are not user-friendly and are tailored to a single technique.
This paper introduces \tool, a plugin for \idea that enables users to generate unit tests with only a few clicks directly within their Integrated Development Environment (IDE). 
Furthermore, \tool also allows users to easily modify and run each generated test and integrate them into the project workflow.
\tool leverages the advances of search-based test generation tools, and it introduces a technique to generate unit tests using Large Language Models (LLMs) by creating a feedback cycle between the IDE and the LLM.
Since \tool is an open-source (\url{https://github.com/JetBrains-Research/TestSpark}), extendable, and well-documented tool, it is possible to add new test generation methods into the plugin with the minimum effort. \new{This paper also explains our future studies related to \tool and our preliminary results.}
\textbf{Demo video:} \url{https://youtu.be/0F4PrxWfiXo}
\end{abstract}



\maketitle

\section{Introduction}
Software testing is essential in the software development process, yet it can be time-consuming and costly \cite{beller2015and}. 
Developers need to manually craft tests that cover various behaviors of their projects.
As a result, numerous studies~\cite{mcminn2011search} propose a range of techniques to automatically generate tests for different testing levels, including unit~\cite{fraser2011evosuite,kex2022,braione2018sushi,derakhshanfar2022basic,lemieux2023codamosa, bareiss2022code, schafer2023empirical}, integration~\cite{derakhshanfar2022generating}, and system level~\cite{arcuri2019restful}. 
These studies confirm that the generated tests not only achieve high code coverage \cite{panichella2018large} but also proved valuable for error detection~\cite{shamshiri2015automatically} and debugging \cite{ceccato2015automatically}.
However, most of these unit test generation tools were primarily designed and used for research studies.
Consequently, openly available test generation tools specialize in one technique and often rely on a command-line interface, making them less user-friendly within development environments like IDEs. 
As a result, users need to interact with each tool separately outside the IDE and, later, integrate tests into their projects manually.

This paper introduces \tool, an open-source, extendable, and well-documented \idea plugin for unit-level test generation of Java programs.
It is designed to ease unit test generation with various techniques.
Currently, \tool supports two technologies: Search-based software test generation (SBST) and Large Language Model (LLM)-based test generation. 
The former is one of the most effective unit test generation techniques~\cite{jahangirova2023sbft}.
The latter has shown potential in helping developers in their software engineering tasks, including software testing~\cite{kang2023large}.
However, a recent study~\cite{tufano2020unit} shows that a large portion ($>50\%$) of tests generated by LLMs are malformed and non-compiling.
\tool introduces an approach (\cref{sec:LLM}) to ensure all tests generated by LLMs are compilable by proposing a feedback loop between the LLM and the IDE. 
\tool is designed to let contributors easily integrate other test generation tools by following our documentation\footnote{\url{https://github.com/JetBrains-Research/TestSpark/blob/main/CONTRIBUTING.md}}. 

Within the IDE, users can seamlessly generate, analyze, modify, and integrate unit tests using \tool.
Our tool is capable of generating tests at various granularity levels, such as for a class, method, or even a single line of code. 
Once tests are generated, \tool offers a visual representation along with a detailed coverage report, encompassing covered lines and killed mutants.
Users have the flexibility to fine-tune \new{(manually or by LLM)} and select the generated tests, while also providing feedback to enhance each test case. 
Finally, users can easily integrate the tests into their projects.

We have released this plugin in the JetBrains Marketplace\footnote{\url{https://plugins.jetbrains.com/plugin/21024-testspark}}.
Despite the recent release, the plugin has garnered positive feedback and more than 1K downloads.
\new{We designed studies to assess the usability of our plugin and the usefulness of the generated tests (explained in Section \ref{sec:eval}). Moreover, the results of our preliminary study show that test generation techniques and features implemented in \tool assist developers in their unit testing tasks.}
\tool is helpful for both software developers and researchers, acting as a bridge between these two communities, all with a singular aim: improving test generation techniques for practical usage. 
Developers can readily use different test generation approaches to assist them in writing unit tests.
Also, researchers can implement their novel strategies within this framework, evaluating them in a development environment and collecting invaluable user feedback.

\section{Related Work}

\textbf{Java Unit Test Generation tools. }
In the past few years, researchers and developers implemented multiple unit test generation tools for Java programs leveraging various techniques (\eg search-based~\cite{fraser2011evosuite,derakhshanfar2022basic}\new{, LLM-based \cite{lemieux2023codamosa, bareiss2022code, schafer2023empirical},} and symbolic execution~\cite{kex2021,kex2022,utbot2022,braione2018sushi}). 
Among these tools, \evosuite~\cite{fraser2011evosuite}, which uses an SBST approach~\cite{panichella2017automated}, outperforms other tools in terms of structural coverage and fault detection~\cite{jahangirova2023sbft}.
%
Most of these Java unit test generation tools use a command-line interface and output their result in one or multiple files (\eg CSV report file and a Java file containing the generated tests).
Hence, they are not fully integrated into IDEs, where developers write tests.
\new{We note that \evosuite also has an \idea plugin \footnote{\url{https://plugins.jetbrains.com/plugin/7985-evosuite-plugin}}, supporting only outdated versions of the IDE (\ie their compatibility range is 15.0 — 2019.3.5).
Moreover, this plugin requires installing \evosuite separately. Also, after test generation, the plugin directly saves the generated tests and their report in a folder.}
\new{In contrast, \tool provides a user-friendly interface in \idea to generate tests for Java code with only one click and visualizes the generated tests and their coverage. Users can interact with each test and use LLMs to enhance the tests generated by each technique.}
\tool is designed to be extendable and supports the integration of multiple test generation tools.
\new{Lastly, \tool is a standalone plugin (\ie does not need any other installation) and supports the latest versions of the \idea.}

\textbf{Java Unit Test Generation plugins in \idea.}
By searching relevant keywords (\eg unit test) in the JetBrains Marketplace, we found five Java unit test generation plugins that are under active development (\ie have at least one yearly release\footnote{Date of query: \textit{Sep 12, 2023}}).
\textit{Kex}
~\cite{kex2021,kex2022} and \textit{UnitTestBot}
~\cite{utbot2022} use symbolic execution for test generation. 
A prior study~\cite{jahangirova2023sbft} confirms that they cannot achieve the structural coverage and fault detection capability of \evosuite. 
In contrast to \tool, these plugins do not provide any visualization for reporting the execution of the generated tests.
\textit{DiffBlue}\footnote{\url{https://plugins.jetbrains.com/plugin/14946-diffblue-cover--ai-for-code}} and \textit{SquareTest}\footnote{\url{https://plugins.jetbrains.com/plugin/10405-squaretest}} are paid plugins that generate Java unit tests.
There is little information about the techniques these plugins use, as their code is not openly available.
These plugins also do not visualize the execution of the generated tests.
Finally, \textit{Chat Unit Test}\footnote{https://plugins.jetbrains.com/plugin/22522-chatunitest} is an LLM-based test generation plugin that uses \openai's \chatgpt to generate tests for a Java class and save the response directly into the code base.
\tool also supports LLM-based test generation; 
\new{however, it can additionally generate tests for individual lines of code. 
Instead of directly saving all tests in a project file, \tool lets users i) analyze each test using the visualized coverage, ii) ask an LLM to apply further enhancement on each test, and iii) select the interesting tests and save them as a new test suite or integrate them into existing test files. \tool is not only an LLM-based test generation framework.
Since \tool has an extendable architecture, new test generators can easily be integrated into this pipeline.
Tests generated by any technique can be improved by LLM upon the user's request. E.g., users can ask LLM to improve the readability of a test case generated by \evosuite.
}

\section{\tool}




\begin{figure}[t]
  \begin{center}
  \includegraphics[width=0.7\linewidth]{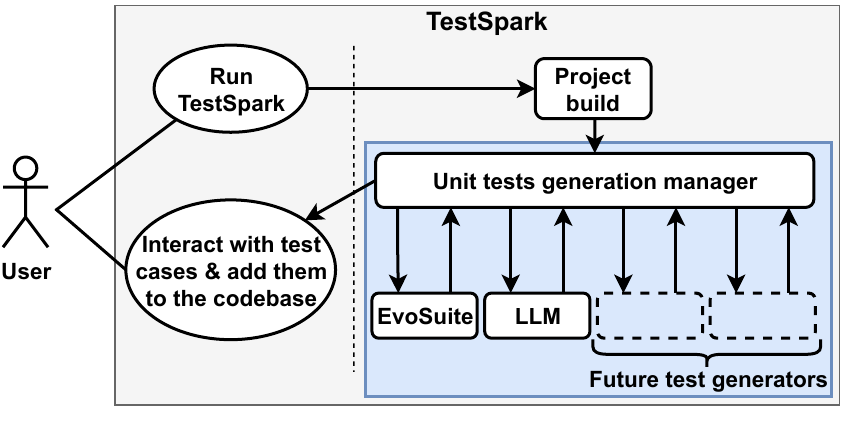}
  \caption{\tool workflow}
  \label{fig:plugin-workflow}
  \end{center}
\end{figure}


\tool is an \idea plugin for generating unit tests that leverages two techniques to bring test generation into the development environment:
SBT (via the state-of-the-art tool \evosuite) and LLM-based test generation (Section \ref{sec:LLM}).
\cref{fig:plugin-workflow} illustrates \tool's workflow.
Users can initiate the test generation process for a unit under test (referred to as \uut) by simply right-clicking on a unit and selecting the \texttt{TestSpark} option.
Upon selecting the desired test generation technique, the process starts.
First, the plugin builds the project within \idea.
This step is crucial as \evosuite requires the compiled code for code instrumentation~\cite{fraser2011evosuite}. 
Similarly, LLM-based test generation requires compilation for test execution and validation based on the model's generative outcome.
Following code compilation, the \textbf{unit test generation manager} employs the chosen technique to initiate the generation process.
Once the tests are collected, they are seamlessly transmitted to the visualization service, where the results are presented to the user (see \cref{sec:visualization}).

\subsection{Interaction with \evosuite}

To integrate \evosuite within \tool, we pursued two main goals: (i) receiving the tests generated by \evosuite and its coverage information inside the plugin instead of reading it from files and (ii) adding a feature to \evosuite that tests individual lines. 
This empowers users to focus the search process on covering specific lines, enhancing the flexibility of the test generator.
This enhanced version of \evosuite is available on GitHub \footnote{\url{https://github.com/ciselab/evosuite/tree/thunderdome}}.
\new{In contrast to the \evosuite \idea plugin, \tool automatically includes the latest release of an enhanced version of \evosuite during the build process.}

To accomplish the first goal, we implemented the \textit{Compact Reporter} module inside \evosuite that generates a serialized report. This report contains all information regarding the tests generated by \evosuite, including test cases and the coverage achieved by each. 

For the second goal, we modified the default test generation algorithm (DynaMOSA~\cite{panichella2017automated}) in \evosuite. DynaMOSA generates tests to cover lines and branches in a selected class or method. This algorithm
\textbf{1)} analyzes the control flow graphs (CFG) of class/method under tests; 
\textbf{2)} it selectively adds branches and lines that are not control-dependent on any other branches into the \textit{active search objectives};
\textbf{3)} it generates a new set of tests and identifies newly covered goals (\ie branches and lines) in the \textit{active search objectives};
\textbf{4)} it saves the tests reaching previously uncovered goals into the archive;
\textbf{5)} it updates the \textit{active search objectives} by removing newly covered goals and adding lines and branches that are control-dependent on these;
\textbf{6)} repeats steps 3 to 6.
In the single-line coverage mode, which we implemented, we made specific modifications to steps 2 and 6.
In these steps, we exclusively add lines and branches that, when covered, provide the test with the opportunity to cover the selected target line (as per the CFG). \new{In other words, this algorithm excludes irrelevant lines and branches to make the search process more focused on covering the target line.}

The plugin generates a command to invoke \texttt{evosuite.jar} for test generation.
This command passes the \uut and the path to the compiled project as parameters.

\subsection{LLM-based test generation}
\label{sec:LLM}

After selecting the preferred LLM platform (\ie the \openai platform and \grazie) and providing the authentication token, users should select the specific LLM model to be used for test generation.
With these configurations in place, users can proceed to request test generation for the \uut.
The plugin then generates a corresponding prompt tailored for the LLM.

\subsubsection{Prompt generation}
The prompt should provide sufficient details for the LLM to generate executable tests.
However, LLMs have a maximum prompt size, so it is necessary to balance the size and the information provided.
To achieve this balance, the prompt generated by \tool should contain the following information: 
(i) the problem description;
(ii) \uut's code; 
(iii) the signatures of methods and objects passed to \uut as input parameters; 
(iv) the polymorphism relations of the classes used by the \uut. \new{Users can modify the prompt template in the settings and add additional information (\eg customized testing objectives or number of test cases).}
Also, the depth of input parameters (used for listing the method signatures) and polymorphism depth are set by the user in \tool settings.
If the prompt generated with the given values exceeds the maximum permitted prompt size, \tool decreases these values iteratively until the prompt is smaller than the permitted size.


\subsubsection{Receiving the LLM response}

\begin{figure}[t]
  \begin{center}
  \includegraphics[width=\linewidth]{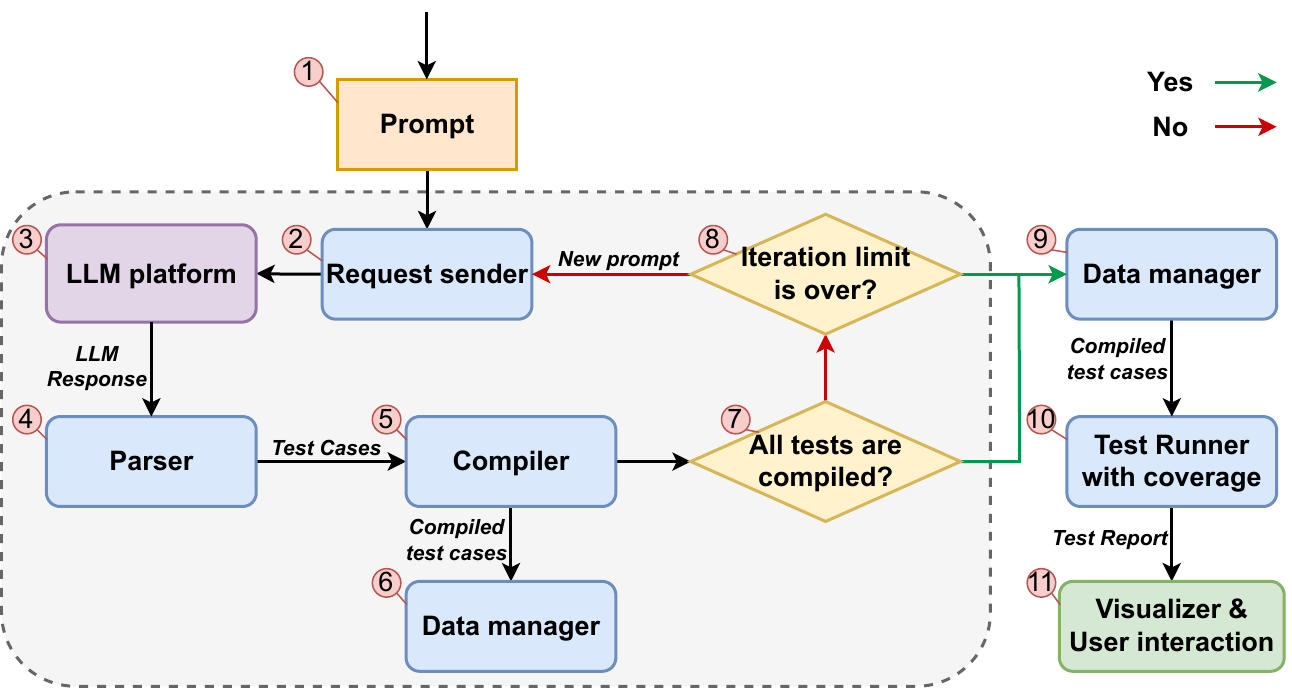}
  \caption{LLM-based test generation}
  \label{fig:LLM-based-test-generation}
  \end{center}
\end{figure}

After prompt generation (\circled{1} in \cref{fig:LLM-based-test-generation}), \tool sends it to the selected LLM (\circled{2}~-~\circled{3}).
Once the LLM's response is received, \tool parses it, extracting the test code written in Java \circled{4}.
Following this, \tool checks each test case individually to identify and save the ones that compile (\circled{5}-\circled{6}).
In case the compilation of the test cases~(\circled{7}) is successful, \tool gathers the code coverage information using \textit{JaCoCo}\footnote{https://www.jacoco.org/jacoco/} (\circled{9}-\circled{11}) and visualizes the results (see \cref{sec:visualization}).
However, if the compilation of the test cases fails and the iteration limit is not reached (\circled{8}), \tool generates a new prompt containing the compilation error and sends a request to the LLM to fix the detected error.
The maximum number of iterations of such a loopback is set by the user in the plugin settings.
If the number of iterations is exhausted, the plugin continues the process with saved (compilable) tests, if any.
Otherwise, it shows an error message and asks to try generating tests for a smaller \uut (\ie method or line).

\subsection{Visualization} \label{sec:visualization}

\begin{figure*}[t]
  \begin{center}
  \includegraphics[width=0.8\linewidth]{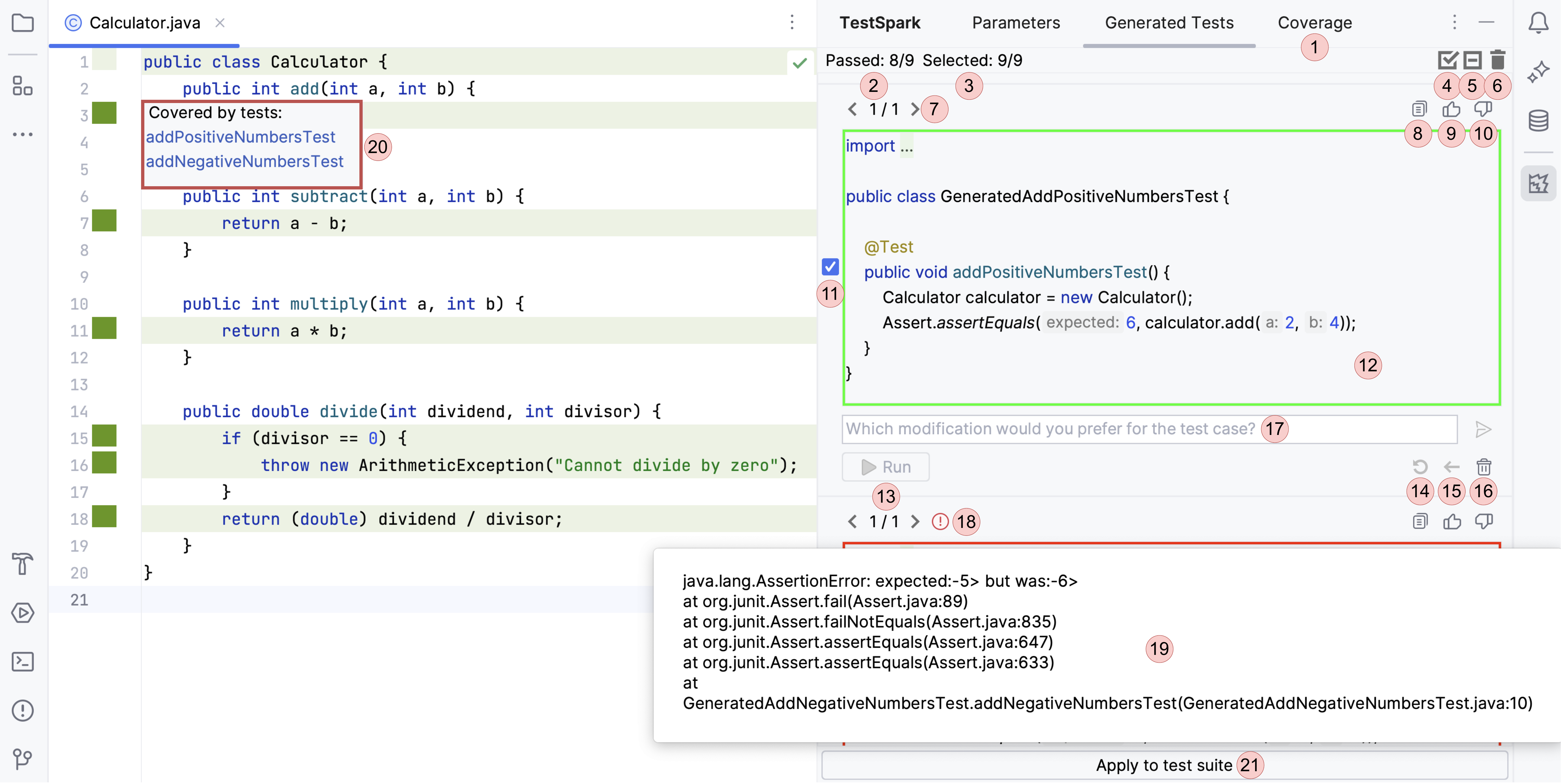}
  \caption{\tool's tests visualization}
  \label{fig:Visualization}
  \end{center}
\end{figure*}

After receiving the result, the user can interact with the test cases in various ways and view their coverage.
\cref{fig:Visualization} illustrates an example of such a result.
The \textit{Coverage} tab shows the structural coverage and mutation score achieved by all generated tests (\circled{1} in \cref{fig:Visualization}).
The table adjusts dynamically, only calculating the statistics for the selected tests.
In the \textit{Generated Test} tab, each generated test is presented.
At the top of this panel, \tool provides the count of passed and selected test cases, along with buttons to select, unselect, and delete all test cases (\circled{2}-\circled{6} in \cref{fig:Visualization}, respectively).

Users have the option to copy, like, dislike, and select each test case (\circled{8}-\circled{11}, respectively). 
Also, the border color of each test block indicates whether it passed (green) or failed (red).
In case of failure, users can hover over the error symbol \circled{18} to view the error message \circled{19}. 
Moreover, the user can modify each test in the code field \circled{12}.
After modification, the "run" \circled{13}, "reset to the initial code" \circled{14}, and "reset to the last run" \circled{15} buttons become available, allowing users to execute the updated test and undo their actions.
Unwanted tests can also be removed \circled{16}.
Additionally, each test case has a text field where users can directly send a modification request to the LLM \circled{17}, \eg adding comments.
This option triggers a new request to the LLM and shows the updated test in a new code window.
The user can switch between the old and updated tests \circled{7}.

Each covered line contains green square on its left. 
Clicking on this square opens a list of test cases that cover this line \circled{20}.
If the user uses \evosuite as generator, \tool shows the list of mutants in this line. 
Clicking on a test case name or a mutant highlights the test cases in the "Generated Tests" that cover or kill it.

Finally, users can integrate the selected/modified tests into their project by clicking the "Apply to test suite" button \circled{21}.
The user can put the tests either in a new or existing file.
In the first case, the user needs to select a folder and then enter the name of the new test class.
In the second case, they can select a Java file to which the new test cases will be added.
In both cases, our plugin adds the required imports, and thereby after integration, the saved/updated test file is compilable and ready to execute.
Moreover, \tool offers the option to change the colors of covered lines for color-blind users.

\new{\section{Evaluation}
\label{sec:eval}
This section explains the designs of our planned studies.
Since our goal for implementing \tool is to aid developers in development and unit testing tasks using test generation,
our research questions are focused on users' interactions with this plugin and how useful they find the test generators and features in \tool: \textbf{RQ1:} \textit{How useful are the generated tests for developers in their day-to-day testing practices?} and \textbf{RQ2:} \textit{How usable is \tool?}

\textbf{Analyzing features usage statistics (FUS):} We are currently implementing collectors to gather anonymized FUS from our 1K+ users. These collectors are designed to answer the following questions: (1) How often do users use each test generation technique? (2) For which \uut do users prefer to generate tests (\eg class, or line)? (3) How often is the test generation successful for each technique? (4) How long does it take to generate tests by each test generation technique? (5) How many tests generated by each technique are integrated into the code base?
(6) which part of the tests are modified by users (\ie test data, method calls, or assertions)? (7) How often do users provide feedback to the LLM (using \circled{17} in \cref{fig:Visualization})? 

\textbf{Questionnaire:}
First, we provide a list of tasks to participants (who have practical experience in unit testing) to use \tool for testing their Java projects. Tasks include generating tests for lines, methods, and classes. Then, we ask participants to analyze the tests and select/integrate the ones that they find useful for their project.
After performing these tasks, the participants should answer a questionnaire about each test generator and feature in \tool. We have conducted a preliminary study based on this questionnaire with two participants (a colleague with unit testing experience and an academic researcher who actively studies software testing). Participants confirmed the usefulness of coverage achieved by generated tests and appreciated the ability to interact with test cases before integrating them into their projects.
Moreover, they noted that generating tests for methods and lines is often more convenient because classes are sometimes too coarse-grained. This confirms the usefulness of the single-line test generation feature. 
}
\section{Conclusion and Future Steps}

In this paper, we have introduced \tool, an open-source unit test generation plugin for \idea. Leveraging cutting-edge techniques, \tool empowers users to effortlessly generate tests for a \uut, all within the familiar IDE environment. 
With a user-friendly interface and a range of features, developers can efficiently manage their testing tasks without switching contexts.
\tool also provides an infrastructure for researchers to implement new test generation approaches and assess these techniques according to user feedback.
%
%
Preliminary observations and feedback indicate the usefulness of \tool in accelerating unit testing tasks. \new{Additionally, we discussed the designs of our planned studies.}
As future steps, we aim to expand \tool's language support. 
%
Also, we aim to combine test generation methods to generate better tests (\eg improve LLM-based technique using SBT).

\begin{acks}
    We thank the students who helped us with the initial prototype of this plugin: Jegor Zelenjak, Martin Mladenov, Kiril Vasilev, Lyuben Todorov, Sergey Datskiv, and Bolek Khodakov.

\end{acks}

\bibliographystyle{ACM-Reference-Format}
\bibliography{bibliography}


\begin{thebibliography}{20}


\ifx \showCODEN    \undefined \def \showCODEN     #1{\unskip}     \fi
\ifx \showDOI      \undefined \def \showDOI       #1{#1}\fi
\ifx \showISBNx    \undefined \def \showISBNx     #1{\unskip}     \fi
\ifx \showISBNxiii \undefined \def \showISBNxiii  #1{\unskip}     \fi
\ifx \showISSN     \undefined \def \showISSN      #1{\unskip}     \fi
\ifx \showLCCN     \undefined \def \showLCCN      #1{\unskip}     \fi
\ifx \shownote     \undefined \def \shownote      #1{#1}          \fi
\ifx \showarticletitle \undefined \def \showarticletitle #1{#1}   \fi
\ifx \showURL      \undefined \def \showURL       {\relax}        \fi
\providecommand\bibfield[2]{#2}
\providecommand\bibinfo[2]{#2}
\providecommand\natexlab[1]{#1}
\providecommand\showeprint[2][]{arXiv:#2}

\bibitem[kex(2021)]%
        {kex2021}
 \bibinfo{year}{2021}\natexlab{}.
\newblock \bibinfo{title}{Kex}.
\newblock
  \bibinfo{howpublished}{\url{https://github.com/vorpal-research/kex/tree/sbst-contest}}.
\newblock


\bibitem[kex(2022)]%
        {kex2022}
 \bibinfo{year}{2022}\natexlab{}.
\newblock \bibinfo{title}{Kex-Reflection}.
\newblock
  \bibinfo{howpublished}{\url{https://github.com/vorpal-research/kex/tree/sbst2022-reflection}}.
\newblock


\bibitem[utb(2022)]%
        {utbot2022}
 \bibinfo{year}{2022}\natexlab{}.
\newblock \bibinfo{title}{UTBot}.
\newblock \bibinfo{howpublished}{\url{https://github.com/UnitTestBot}}.
\newblock


\bibitem[Arcuri(2019)]%
        {arcuri2019restful}
\bibfield{author}{\bibinfo{person}{Andrea Arcuri}.}
  \bibinfo{year}{2019}\natexlab{}.
\newblock \showarticletitle{RESTful API automated test case generation with
  EvoMaster}.
\newblock \bibinfo{journal}{\emph{ACM Transactions on Soft. Eng. and
  Methodology (TOSEM)}} \bibinfo{volume}{28}, \bibinfo{number}{1}
  (\bibinfo{year}{2019}), \bibinfo{pages}{1--37}.
\newblock


\bibitem[Barei{\ss} et~al\mbox{.}(2022)]%
        {bareiss2022code}
\bibfield{author}{\bibinfo{person}{Patrick Barei{\ss}},
  \bibinfo{person}{Beatriz Souza}, \bibinfo{person}{Marcelo d’Amorim}, {and}
  \bibinfo{person}{Michael Pradel}.} \bibinfo{year}{2022}\natexlab{}.
\newblock \bibinfo{title}{Code Generation Tools (Almost) for Free? A Study of
  Few-Shot, Pre-Trained Language Models on Code. CoRR abs/2206.01335 (2022)}.
\newblock
\newblock


\bibitem[Beller et~al\mbox{.}(2015)]%
        {beller2015and}
\bibfield{author}{\bibinfo{person}{Moritz Beller}, \bibinfo{person}{Georgios
  Gousios}, \bibinfo{person}{Annibale Panichella}, {and} \bibinfo{person}{Andy
  Zaidman}.} \bibinfo{year}{2015}\natexlab{}.
\newblock \showarticletitle{When, how, and why developers (do not) test in
  their IDEs}. In \bibinfo{booktitle}{\emph{Proceedings of the 2015 10th Joint
  Meeting on Foundations of Software Engineering}}. \bibinfo{pages}{179--190}.
\newblock


\bibitem[Braione et~al\mbox{.}(2018)]%
        {braione2018sushi}
\bibfield{author}{\bibinfo{person}{Pietro Braione}, \bibinfo{person}{Giovanni
  Denaro}, \bibinfo{person}{Andrea Mattavelli}, {and} \bibinfo{person}{Mauro
  Pezz{\`e}}.} \bibinfo{year}{2018}\natexlab{}.
\newblock \showarticletitle{SUSHI: a test generator for programs with complex
  structured inputs}. In \bibinfo{booktitle}{\emph{Proceedings of the 40th
  International Conference on Software Engineering: Companion Proceeedings}}.
  \bibinfo{pages}{21--24}.
\newblock


\bibitem[Ceccato et~al\mbox{.}(2015)]%
        {ceccato2015automatically}
\bibfield{author}{\bibinfo{person}{Mariano Ceccato},
  \bibinfo{person}{Alessandro Marchetto}, \bibinfo{person}{Leonardo Mariani},
  \bibinfo{person}{Cu~D Nguyen}, {and} \bibinfo{person}{Paolo Tonella}.}
  \bibinfo{year}{2015}\natexlab{}.
\newblock \showarticletitle{Do automatically generated test cases make
  debugging easier? an experimental assessment of debugging effectiveness and
  efficiency}.
\newblock \bibinfo{journal}{\emph{ACM Transactions on Soft. Eng. and
  Methodology (TOSEM)}} \bibinfo{volume}{25}, \bibinfo{number}{1}
  (\bibinfo{year}{2015}), \bibinfo{pages}{1--38}.
\newblock


\bibitem[Derakhshanfar and Devroey(2022)]%
        {derakhshanfar2022basic}
\bibfield{author}{\bibinfo{person}{Pouria Derakhshanfar} {and}
  \bibinfo{person}{Xavier Devroey}.} \bibinfo{year}{2022}\natexlab{}.
\newblock \showarticletitle{Basic block coverage for unit test generation at
  the SBST 2022 tool competition}. In \bibinfo{booktitle}{\emph{Proceedings of
  the 15th Workshop on Search-Based Software Testing}}.
  \bibinfo{pages}{37--38}.
\newblock


\bibitem[Derakhshanfar et~al\mbox{.}(2022)]%
        {derakhshanfar2022generating}
\bibfield{author}{\bibinfo{person}{Pouria Derakhshanfar},
  \bibinfo{person}{Xavier Devroey}, \bibinfo{person}{Annibale Panichella},
  \bibinfo{person}{Andy Zaidman}, {and} \bibinfo{person}{Arie van Deursen}.}
  \bibinfo{year}{2022}\natexlab{}.
\newblock \showarticletitle{Generating Class-Level Integration Tests Using Call
  Site Information}.
\newblock \bibinfo{journal}{\emph{IEEE Transactions on Software Engineering}}
  \bibinfo{volume}{49}, \bibinfo{number}{4} (\bibinfo{year}{2022}),
  \bibinfo{pages}{2069--2087}.
\newblock


\bibitem[Fraser and Arcuri(2011)]%
        {fraser2011evosuite}
\bibfield{author}{\bibinfo{person}{Gordon Fraser} {and} \bibinfo{person}{Andrea
  Arcuri}.} \bibinfo{year}{2011}\natexlab{}.
\newblock \showarticletitle{Evosuite: automatic test suite generation for
  object-oriented software}. In \bibinfo{booktitle}{\emph{Proceedings of the
  19th ACM SIGSOFT symposium and the 13th European conference on Foundations of
  software engineering}}. \bibinfo{pages}{416--419}.
\newblock


\bibitem[Jahangirova and Terragni(2023)]%
        {jahangirova2023sbft}
\bibfield{author}{\bibinfo{person}{Gunel Jahangirova} {and}
  \bibinfo{person}{Valerio Terragni}.} \bibinfo{year}{2023}\natexlab{}.
\newblock \showarticletitle{SBFT tool competition 2023-Java test case
  generation track}. In \bibinfo{booktitle}{\emph{2023 IEEE/ACM International
  Workshop on Search-Based and Fuzz Testing (SBFT)}}. IEEE,
  \bibinfo{pages}{61--64}.
\newblock


\bibitem[Kang et~al\mbox{.}(2023)]%
        {kang2023large}
\bibfield{author}{\bibinfo{person}{Sungmin Kang}, \bibinfo{person}{Juyeon
  Yoon}, {and} \bibinfo{person}{Shin Yoo}.} \bibinfo{year}{2023}\natexlab{}.
\newblock \showarticletitle{Large language models are few-shot testers:
  Exploring llm-based general bug reproduction}. In
  \bibinfo{booktitle}{\emph{2023 IEEE/ACM 45th International Conference on
  Software Engineering (ICSE)}}. IEEE, \bibinfo{pages}{2312--2323}.
\newblock


\bibitem[Lemieux et~al\mbox{.}(2023)]%
        {lemieux2023codamosa}
\bibfield{author}{\bibinfo{person}{Caroline Lemieux},
  \bibinfo{person}{Jeevana~Priya Inala}, \bibinfo{person}{Shuvendu~K Lahiri},
  {and} \bibinfo{person}{Siddhartha Sen}.} \bibinfo{year}{2023}\natexlab{}.
\newblock \showarticletitle{CODAMOSA: Escaping coverage plateaus in test
  generation with pre-trained large language models}. In
  \bibinfo{booktitle}{\emph{International conference on software engineering
  (ICSE)}}.
\newblock


\bibitem[McMinn(2011)]%
        {mcminn2011search}
\bibfield{author}{\bibinfo{person}{Phil McMinn}.}
  \bibinfo{year}{2011}\natexlab{}.
\newblock \showarticletitle{Search-based software testing: Past, present and
  future}. In \bibinfo{booktitle}{\emph{2011 IEEE Fourth International
  Conference on Software Testing, Verification and Validation Workshops}}.
  IEEE, \bibinfo{pages}{153--163}.
\newblock


\bibitem[Panichella et~al\mbox{.}(2017)]%
        {panichella2017automated}
\bibfield{author}{\bibinfo{person}{Annibale Panichella},
  \bibinfo{person}{Fitsum~Meshesha Kifetew}, {and} \bibinfo{person}{Paolo
  Tonella}.} \bibinfo{year}{2017}\natexlab{}.
\newblock \showarticletitle{Automated test case generation as a many-objective
  optimisation problem with dynamic selection of the targets}.
\newblock \bibinfo{journal}{\emph{IEEE Transactions on Software Engineering}}
  \bibinfo{volume}{44}, \bibinfo{number}{2} (\bibinfo{year}{2017}),
  \bibinfo{pages}{122--158}.
\newblock


\bibitem[Panichella et~al\mbox{.}(2018)]%
        {panichella2018large}
\bibfield{author}{\bibinfo{person}{Annibale Panichella},
  \bibinfo{person}{Fitsum~Meshesha Kifetew}, {and} \bibinfo{person}{Paolo
  Tonella}.} \bibinfo{year}{2018}\natexlab{}.
\newblock \showarticletitle{A large scale empirical comparison of
  state-of-the-art search-based test case generators}.
\newblock \bibinfo{journal}{\emph{Information and Software Technology}}
  \bibinfo{volume}{104} (\bibinfo{year}{2018}), \bibinfo{pages}{236--256}.
\newblock


\bibitem[Sch{\"a}fer et~al\mbox{.}(2023)]%
        {schafer2023empirical}
\bibfield{author}{\bibinfo{person}{Max Sch{\"a}fer}, \bibinfo{person}{Sarah
  Nadi}, \bibinfo{person}{Aryaz Eghbali}, {and} \bibinfo{person}{Frank Tip}.}
  \bibinfo{year}{2023}\natexlab{}.
\newblock \showarticletitle{An Empirical Evaluation of Using Large Language
  Models for Automated Unit Test Generation}.
\newblock \bibinfo{journal}{\emph{IEEE Transactions on Software Engineering}}
  (\bibinfo{year}{2023}).
\newblock


\bibitem[Shamshiri et~al\mbox{.}(2015)]%
        {shamshiri2015automatically}
\bibfield{author}{\bibinfo{person}{Sina Shamshiri}, \bibinfo{person}{Ren{\'e}
  Just}, \bibinfo{person}{Jos{\'e}~Miguel Rojas}, \bibinfo{person}{Gordon
  Fraser}, \bibinfo{person}{Phil McMinn}, {and} \bibinfo{person}{Andrea
  Arcuri}.} \bibinfo{year}{2015}\natexlab{}.
\newblock \showarticletitle{Do automatically generated unit tests find real
  faults? an empirical study of effectiveness and challenges (t)}. In
  \bibinfo{booktitle}{\emph{2015 30th IEEE/ACM Int. Conference on Automated
  Software Engineering (ASE)}}. IEEE, \bibinfo{pages}{201--211}.
\newblock


\bibitem[Tufano et~al\mbox{.}(2020)]%
        {tufano2020unit}
\bibfield{author}{\bibinfo{person}{Michele Tufano}, \bibinfo{person}{Dawn
  Drain}, \bibinfo{person}{Alexey Svyatkovskiy}, \bibinfo{person}{Shao~Kun
  Deng}, {and} \bibinfo{person}{Neel Sundaresan}.}
  \bibinfo{year}{2020}\natexlab{}.
\newblock \showarticletitle{Unit test case generation with transformers and
  focal context}.
\newblock \bibinfo{journal}{\emph{arXiv preprint arXiv:2009.05617}}
  (\bibinfo{year}{2020}).
\newblock


\end{thebibliography}

\end{document}